\documentclass[fleqn,10pt]{SelfArx} 

\definecolor{color1}{RGB}{70,70,90} 
\definecolor{color2}{RGB}{0,20,20} 

\newlength{\tocsep} 
\usepackage{lipsum} 
\usepackage{inputenc}
\usepackage{listings}
\usepackage{color}
\usepackage{graphicx}
\usepackage{float}

\definecolor{sh_comment}{rgb}{0.12, 0.38, 0.18 } 
\definecolor{sh_keyword}{rgb}{0.37, 0.08, 0.25}  
\definecolor{sh_string}{rgb}{0.06, 0.10, 0.98} 

\JournalInfo{LabGTP} 
\Archive{ICAR-CNR} 

\PaperTitle{The MetaboX library: building metabolic networks from KEGG database} 

\Authors{F. Maiorano\textsuperscript{1}*, L. Ambrosino\textsuperscript{2}, M.R. Guarracino\textsuperscript{1}} 
\affiliation{\textsuperscript{1}\textit{Laboratory for Genomics, Transcriptomics and Proteomics, Institute for High-Performance Computing and Networking, National Research Council, Via P. Castellino 111, 80131 (NA), Italy}} 
\affiliation{\textsuperscript{2}\textit{Dept. of Agricultural Sciences, University of Naples Federico II, Via Universit\`a 100, 80055 Portici (NA), Italy}} 
\affiliation{*\textbf{Corresponding author}: francesco.maiorano@na.icar.cnr.it} 

\newcommand{\keywordname}{Keywords} 

\Abstract{
In many key applications of metabolomics, such as toxicology or nutrigenomics, it is of interest to profile and detect changes in metabolic processes, usually represented in the form of pathways. As an alternative, a broader point of view would enable investigators to better understand the relations between entities that exist in different processes. Therefore, relating a possible perturbation to several known processes represents a new approach to this field of study. We propose to use a network representation of metabolism in terms of reactants, enzyme and metabolite. To model these systems it is possible to describe both reactions and relations among enzymes and metabolites. In this way, analysis of the impact of changes in some metabolites or enzymes on different processes are easier to understand, detect and predict. 

\textbf{Results}: We release the MetaboX library, an open source PHP framework for developing metabolic networks from a set of compounds. This library uses data stored in \textit{Kyoto Encyclopedia for Genes and Genomes} (KEGG) database using its RESTful \textit{Application Programming Interfaces} (APIs), and methods to enhance manipulation of the information returned from KEGG webservice. The MetaboX library includes methods to extract information about a resource of interest (e.g. metabolite, reaction and enzyme) and to build reactants networks, bipartite enzyme-metabolite and unipartite enzyme networks. These networks can be exported in different formats for data visualization with standard tools. As a case study, the networks built from a subset of the Glycolysis pathway are described and discussed.

\textbf{Conclusions}: The advantages of using such a library imply the ability to model complex systems with few starting information represented by a collection of metabolites KEGG IDs.
}

\begin{document}

\flushbottom 

\maketitle 


\thispagestyle{empty} 


\section{Introduction}
In metabolomics applications it is often of interest to model relationships and interactions among compounds and enzymes, such as protein-protein interactions, metabolite pathways or pathway flows. 
One of the main challenges in metabolomics is to model these interactions in the form of networks\\\cite{MNM,NBF}. Such networks make it easier to understand the topological and functional structure of molecules and their interactions. Furthermore, once the network has been built, various statistics can be obtained for characterization and comparison.  Indeed, a network represents a convenient way to model objects and their relationships as complex systems.

Modeling a network of metabolites provides several ways to further analyze types of interactions, to understand the role of each metabolite in a particular pathway and to detect changes. The problem we address is to build \textit{reaction}, \textit{unipartite enzymes} and \textit{bipartite enzyme-metabolite} networks, starting from a list of metabolites and information on metabolism gathered from a database. These networks models were used in other research studies in order to identify so-called reporter metabolites \cite{KiranRaos}. In a metabolite reaction network, two metabolites are connected if they are known to react in the same reaction. In a unipartite enzyme network, two enzymes are connected if they share at least one metabolite in the reactions they catalyze. In a bipartite enzyme-compound network, each enzyme is connected to every metabolite that is present in the reactions it catalyzes.

There are several publicly available databases that store and distribute information on molecular compounds providing different access methods. Among these we cite MetaCyc \cite{metacyc}, EcoCyc \cite{ecocyc}, HMDB \cite{hmdb}, Lipid Maps \cite{lmsd}, BioCyc \cite{biocyc}, Reactome \cite{reactome}, PubChem \cite{pubchem}, Chebi \cite{chebi}, ChemSpider \cite{chemspider}, Meltin \cite{meltin}, IIMDB \cite{iimdb}, and KEGG \cite{KEGG}.
It is out of the scope of this paper to describe the characteristics of all these databases, and we focus our attention on the latter. KEGG is a database containing the largest collection of metabolites, as well as enzymes, reactions and other information\cite{dbcomp}. It is possible to query the database with a web interface using one compound, and obtain information on the reactions in which it is involved, the stoichiometric equations, enzymes that catalyze the reaction, and metabolic pathways in which these reactions are involved. Its website graphically displays the stored pathways, but there is no functionality to build networks with a custom topology. To overcome these difficulties, some software exist, and partially solve these problems.

KEGGgraph \cite{kegggraph} represents an interface between KEGG pathways and graph objects. It parses KGML (KEGG XML) files into graph models. This tool only provides modeling for KEGG pathways and it is only available for R.
MetaboAnalyst \cite{metaboanalyst} provides a web-based analytical pipeline for metabolomic studies. Its web interface can be used to load data as a list, for statistical analysis, as well as pathway analysis. When queried with a list of compounds, it returns information on pathways taken from KEGG but, for reasons related to XML representation of KEGG pathways, information concerning reactions and substrates are partially lost. Therefore, the resulting metabolic network is often disconnected and it does not represent a good model for graph analysis. The source code is not available, neither it provides APIs of any form.

INMEX \cite{inmex}, introduces an integrative meta-analysis of expression data and a web-based tool to support meta-analysis. It provides a web interface to perform complex operations step-by-step. While it supports custom data processing, annotation and visualization, it does not provide any APIs to extend core functionalities and it cannot be deployed in a custom environment. MetaboLyzer \cite{metabolyzer} implements a  workflow for statistical analysis of metabolomics data. It aims at both simplifying  analysis for investigators who are new to metabolomics, and providing the flexibility to conduct sophisticated analysis to experienced investigators. It uses KEGG, HMDB, Lipid Maps, and BioCyc for putative ion identification. However, it is specifically suited for analysis of post-processed liquid chromatography-mass spectrometry (LC–MS)-based metabolomic data sets. Finally, a tool that implements network construction is MetaboNetworks \cite{metabonetworks} which builds the networks using main reaction pairs and provide analyses for specific organisms.

Although these software give the possibility to model a new network starting from a list of compounds, providing relevant tools for statistical and functional analyses, they miss the cabability to programmatically query metabolomics resources in order to develop novel applications and the software development choices made them suited for very specific environments raising difficulties to use them in production.

With the aim to fill that gap, we introduce the MetaboX library which is a framework that enables investigators to extract information that is not visible at a first glance in KEGG. In fact, it is possible to retrieve many information available in the database with just a collection of KEGG IDs and then connect the gathered data in ways KEGG does not provide.
On the other hand, it is possible to programmatically query the database, obtaining such information in the form of flat files. For large lists of input nodes it can be very difficult to manually gather information of interest to build a network, as well as other metadata useful to get a complete understanding of the biological system.

With respect to the tools presented above, MetaboX provides a framework to model custom network layouts from a list of input nodes. Based on the nature of input nodes, we provide a set of classes to gather related information and programmatically build a network. The design of the MetaboX library is suited for web production environments, in fact it can be embedded in a custom webservice as it is released under the AGPL license. Therefore, the MetaboX library is an open source framework that aims to get a growing community of researchers and developers to support metabolomic analysis.

With the MetaboX library, developers are able to model a network in different ways using the available methods to create a custom network layout that meets their needs.
The library design is modular, with the aim to give developers the ability to implement different types of network builders from lists of compounds. Thus, gathering information and detecting interactions programmatically represents a benefit when working with large lists of metabolites. In the network construction process, MetaboX handles the following steps:

\begin{itemize}
\item \emph{connect} to the resource provider database using the PHP {\tt libcurl} library 
\item \emph{query} a resource provider using methods to retrieve nodes and interactions. As KEGG does not provide a structured query response, we built a translation layer to extract information from flat files. 
\item \emph{extract} requested resource attributes from returned data, parsing and storing them. This task is achieved using regular expressions.
\item \emph{cache} resource attributes to file using a convenient data structure for serialization and for sharing and processing purposes such as JSON. 
\item \emph{build} a consistent data structure with information about requested resources using all previous steps, in order to build a network from collected data. Results consist of a weighted edgelist and a list of network nodes. It also includes specific resource information.
\end{itemize}

In the \textit{connect} step, MetaboX currently supports HTTP, HTTPS, FTP, and ldap protocols. It also supports HTTPS certificates, POST, PUT and FTP uploading, which are natively available in the \textit{curl} library. 
It is also possible to \textit{query} the KEGG database with a list of resources of interest and then parse the response to firstly separate information about each one and then extract specific data. In the \textit{extract} step downloaded data are parsed to produce new files ready for next steps.
We locally \textit{cache} data to load resource information every time it is requested again by a new process. We design the caching system for MetaboX to speed up computation and to produce a sustainable amount of requests to the resource provider system.

It is possible to invalidate the cache in order to reload updated data, and to manually delete the cache so that the library can update it. Finally, the \emph{build} step is intended to put together all gathered information and output the resulting network. Every build method connects two nodes differently in each network model, that is \emph{Reactant Graph} implementation of the build method is different from both \emph{Enzyme Unipartite Graph} and \emph{Enzyme Bipartite Graph}.
In the following section, we report the implementation of the MetaboX library, detailing how it provides easy access to KEGG database and data manipulation. We explain how to use the library to build the different networks proposed and how to export the result for visualization and analysis with external tools like Cytoscape \cite{cytoscape}, which we use to render the figures presented in this paper. Then, we provide a case study and discuss the results. Finally, we conclude providing details on future work directions and open problems.

\section{Implementation}
At the moment of this writing, KEGG only offers a RESTful API interface thus MetaboX is designed to query these in an appropriate manner. KEGG used to expose SOAP APIs to standard software but these were suppressed on 31st december 2012 and the toolbox does not work anymore\\(https://www.biocatalogue.org/announcements/37).

KEGG returns plain text upon web-service calls, thus making it necessary to parse results and arrange them in a data structure. We query KEGG multiple times and store the gathered information to file. The file format we use is JSON which is a lightweight data-interchange format, human-readable and writable. JSON is a text format that is completely language independent but uses conventions that are familiar to C-family programmers. These properties make JSON an ideal data-interchange language.

To limit requests to KEGG, the MetaboX library loads previously processed resources from local storage, if they are available.
A sample request for a resource in KEGG can be achieved using the following url:\\http://rest.kegg.jp/$<$operation$>$/$<$argument$>$
For instance, to retrieve information about metabolite C01290, we use\\http://rest.kegg.jp/get/cpd:C01290. We provide details about the implementation in the following sections.

\subsection{Network Construction}
We deal with compounds, reactions and pathways. To handle such a variety of entities the MetaboX library contains classes to instantiate them. \emph{AbstractResourceLoader} is an abstract class that provides methods needed to load an entity. To model an entity with a new class, this has to extend the abstract class and implement the abstract \emph{load} method. When an entity is instatiated, this method first checks for existing records in the cache. If the requested entity has not been processed previously, a new file is built upon KEGG response. This pattern is used to load metabolites, reactions, pathways and enzymes. We provide several helper methods to extract information about resources from plain text using regular expressions. When the entity has been successfully processed, we serialize it to file for further reference.

The attributes that define a metabolite are: \emph{id}, \emph{formula}, \emph{exact mass}, \emph{molecular weight}, \emph{reaction list}, \emph{pathway list}, \emph{enzyme list}. Lists of other entities of interest that are related to a metabolite, such as reactions, pathways and enzymes, are loaded with different API call. For instance, if the \emph{load} of C01290 returns a list of 10 reactions\\(http://rest.kegg.jp/get/cpd:C01290), we use a RESTful url (http://rest.kegg.jp/get/rn:R\_id) to instantiate each of these reactions. It is possible to query KEGG RESTful APIs using collections of metabolites, reactions, enzymes or pathways. Using this capability, we designed the MetaboX library to construct queries splitting the input collection in chunks of 10 items (as this is the maximum chunk size KEGG supports).
For reactions, we collect \emph{id}, \emph{name}, \emph{definition}, \emph{equation}, \emph{enzymes} and \emph{pathways}. For data manipulation purposes and to conveniently organize reaction information about input metabolites, we process reaction equations separating reactants from products in a data structure.

Cache directories can be set in a configuration file. Each resource is stored in a dedicated resource directory and files are named after resource id (e.g. \{resource\}/\{resource\_id\}.json would result in compound/C00002.json). The configuration file 'config.ini' is divided in sections and it is possible to specify storage directories for entities (e.g. config-$>$directory-$>$compound or config-$>$directory-$>$reaction) as well as KEGG API urls (e.g. config-$>$url-$>$compound or\\
config-$>$url-$>$reaction). This approach is helpful if the entities become available in different urls or from another resource provider.

In the MetaboX library we provide an interface to build several networks. \emph{AbstractGraphBuilder} is an abstract class that defines the general structure of the resulting network. Specific network builder classes must implement the abstract \emph{build} method provided in the abstract builder which takes one optional parameter. This is a list of metabolites out of which a sub network has to be built. To create a new type of network, a builder class should provide the construction of a network involving input metabolites and others involved in common reactions, or other entities, such as enzymes. If the optional parameter is specified, the builder method should create a network with set of nodes given by input parameter. When the network-construction process is completed, \emph{getGlobalGraph} and \emph{getSubGraph} methods return a multidimensional array containing the list of nodes, a weighted edgelist, where the weight represents the number of times a reaction has been found, and the list of connected and not connected nodes, in the case of a sub network.

\subsection{Reactants Network}
A network of reactants $G = (V, E)$ is an undirected graph where each node represents a metabolite and two given nodes $A$ and $B$ in $V$ interact with each other only if there is at least one reaction equation where $A$ and $B$ are involved as reactants.
\emph{ReactantsGraph} class builds a network out of a list of metabolites. To achieve this task, we first gather metabolites and reactions data from KEGG (Listing \ref{lst:resourcesloader}). We create a list of reactions that involve input metabolites and pass it to the class. In this case, the \emph{build} method cycles through the list of reactions and, for each one, the list of substrates is extracted. We then connect each substrate to one another and when all direct network interactions have been built, we produce a weighted edgelist. Such edgelist represents a network including input compounds and all other compounds involved in processed reactions.

We also save a weighted list of interactions that only include input compounds, this resulting in a smaller network which can be seen as a sub network of the global weighted interaction list. As shown in Fig. \ref{fig:global_network}, the sub network is embedded in the global one. A builder class exposes methods to compute results and pass them to the graph writer classes in order to produce a file format that is suitable to the needs of further analysis, such as SIF and XML.

The modeling of this class of networks allows to detect which compounds are directly connected, being reagents of the same reactions. It highlights what are the highly connected hubs in a network made up of the collection of metabolites under analysis. This information is useful for planning metabolic engineering strategies. It is clear that if we wish to modify a node of this type of network, it is crucial to know what are other reactants to be considered, so that the change can effectively impact on the metabolic system of the studied biological organism.

\subsection{Bipartite Enzyme-Metabolite Network}
A network of enzymes and metabolites is a bipartite undirected graph $Z = (U, V, E)$ with set of nodes $U$ representing metabolites and $V$ representing enzymes. A metabolite node is connected to all the enzymes nodes that catalyze a reaction involving that metabolite, and an enzyme node is connected to all the metabolites that take part in the corresponding reaction. That is, if an enzyme $F$ in $V$ catalyzes a reaction where a metabolite $M$ in $U$ is a substrate, then an interaction between $F$ and $M$ exists in the network.

We achieve this task using \emph{EnzymeBipartiteGraph} class which parameters are: a metabolite collection, an enzyme collection and a reaction collection. We cycle through the list of metabolites and select the related enzymes. We search current metabolite $M$ in the substrates of the reaction catalyzed by enzyme $F$. If we have a match, we connect nodes $F$ and $M$.

\begin{figure}[h!]
\centering
\includegraphics[scale=0.3]{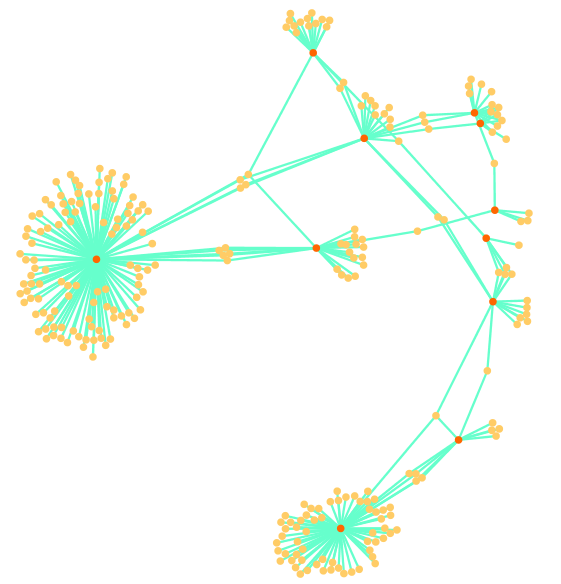}
\caption{Enzyme-metabolite bipartite network: 342 nodes and 393 interactions. Darker nodes represent metabolites.}
\label{fig:ec_bipartite}
\end{figure}

An enzymes network, both unipartite and bipartite, provides a kind of visualization that highlights some aspects that are not observable by a reactants network. If we are analyzing different time conditions with different concentration levels of some compounds, for instance, this class of networks would quickly identify which nodes are most affected, restricting the area of interest to the enzyme directly susceptible to a particular condition. Therefore, the construction of this type of graphs can help highlight changes in the enzymatic expression levels or to detect enzymes with structural or functional defects due to particular conditions of stress. An example of such a network is shown in Fig. \ref{fig:ec_bipartite}.

\subsection{Unipartite Enzymes Network}
\begin{figure}[h!]
\centering
\includegraphics[scale=0.2]{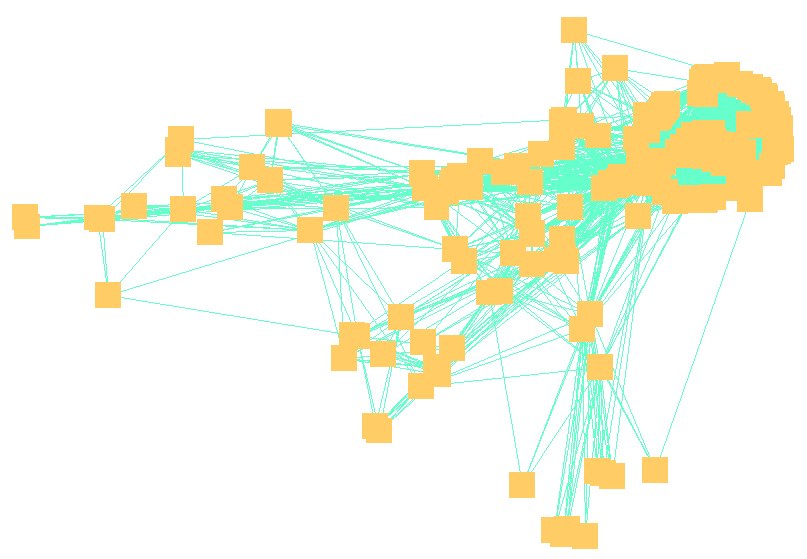}
\caption{Enzymes unipartite network: 297 nodes and 7705 interactions.}
\label{fig:ec_unipartite}
\end{figure}

A unipartite network of enzymes is an undirected graph $G = (V, E)$ where nodes represent enzymes and two enzymes sharing a common compound in the corresponding reactions are connected to each other. The class used to model such a network is \emph{EnzymeUnipartiteGraph}. This builder class is instantiated with a list of enzymes and a list of reactions. These lists are created collecting all reactions and enzymes that involve input metabolites (Listing \ref{lst:enzymeloader}). For each enzyme in the collection, we load data of the reaction catalyzed and select all substrates. We cycle through the enzyme collection comparing the current enzyme substrates to all others. Given two enzymes $T$ and $S$ in $V$, we connect them if the intersection between substrates in $T$ and substrates in $S$ is not empty. An example of such a network is shown in Fig. \ref{fig:ec_unipartite}.

\subsection{Data Export}
In the MetaboX library, there are two classes that can be used to export the constructed network in other formats. As for the other components, a \emph{AbstractGraphWriter} is an abstract class that exposes an abstract \emph{write} method. The class constructor takes one parameter, that is a multidimensional array containing the node list and the weighted edgelist of the network. This can be set using \emph{getGlobalGraph} or \emph{getSubGraph} to export respectively a network or a sub network. The \emph{write} method has two parameters: the name of the file to be written and the data that needs to be exported. If the output needs to be prepared or modified somehow, it is possible to call \emph{prepareOutput} within the \emph{write} method. This is the case of \emph{CytoscapeGraphWriter} class where interactions are converted to string and then written to file. To work with the D3JS (http://d3js.org) visualization library as well as D3py (https://github.com/mikedewar/d3py) or NetworkX \cite{networkx} (http://networkx.github.io), the MetaboX library provides several classes to export a network in one of the formats accepted by other analysis tools. For instance, a \emph{D3JSGraphWriter} converts the network to JSON and writes it to file.

\section{Results and Discussion}
\label{sec:resdisc}
In order to test MetaboX, we built a network starting from a set of eleven compounds, listed below:
\begin{itemize}
\item Glucose (C00031)
\item Glucose 6-phosphate (C00668)
\item Fructose 6-phosphate (C05345)
\item Fructose 1,6-bisphosphate (C05378)
\item Dihydroxyacetone phosphate (C00111)
\item Glyceraldehyde 3-phosphate (C00118)
\item 1,3-bisphosphoglycerate (C00236)
\item 3-phosphoglycerate (C00197)
\item 2-phosphoglycerate (C00631)
\item Phosphoenolpyruvate (C00074)
\item Pyruvate (C00022)
\end{itemize}

\begin{figure}[h!]
\centering
\includegraphics[scale=0.2]{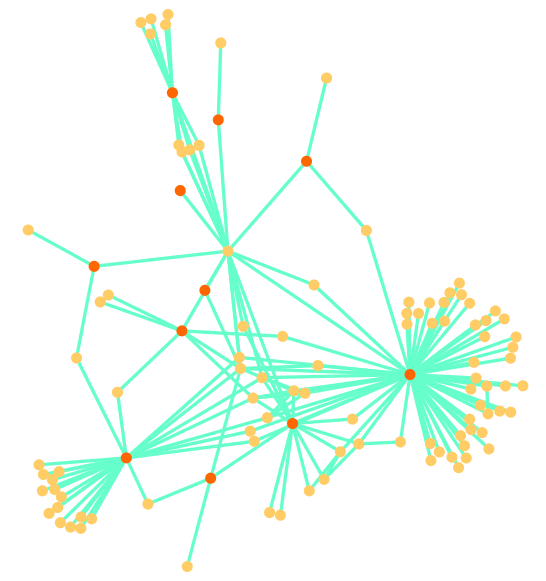}
\caption{A network of reactants obtained from the 11 input metabolites (darker nodes) selected from glycolysis pathway. This network represents the potential interactions among glycolysis metabolites and any other metabolite involved in at least one reaction involving input metabolites. This network shows 108 nodes and 151 edges.}
\label{fig:global_network}
\end{figure}

These compounds belong to glycolysis, the metabolic pathway that leads, starting from glucose, to the production of pyruvic acid through several reactions. Fig. \ref{fig:global_network} shows the network built out of all reactions involving input metabolites. This results in a network that includes input metabolites as well as others related to them. Here we found 151 interactions between 108 metabolites, including all the input metabolites. An interaction in this graph means to be reactants of the same reactions, therefore each node will be directly affected by a decrease in the concentration of one of its neighbors, but not by an increase. In this second case, in fact, there would be an excess of one of the two reactants. This information can be useful if we want, for instance, to plan a change in the levels of a compound starting from other compounds already known. Glucose and Pyruvic acid, as well as representing the start and the end of the glycolysis, are also hubs, namely highly connected nodes, in the network shown in Fig. \ref{fig:global_network}. Glucose is a widely used carbohydrate in organisms, and cells use glucose as a secondary source of energy and a metabolic intermediate. Pyruvic acid is known to be a key intersection in several metabolic pathways. It is clear how a change in these highly connected nodes may severely affect the metabolism of an organism.

\begin{figure}[h!]
\centering
\includegraphics[scale=0.3]{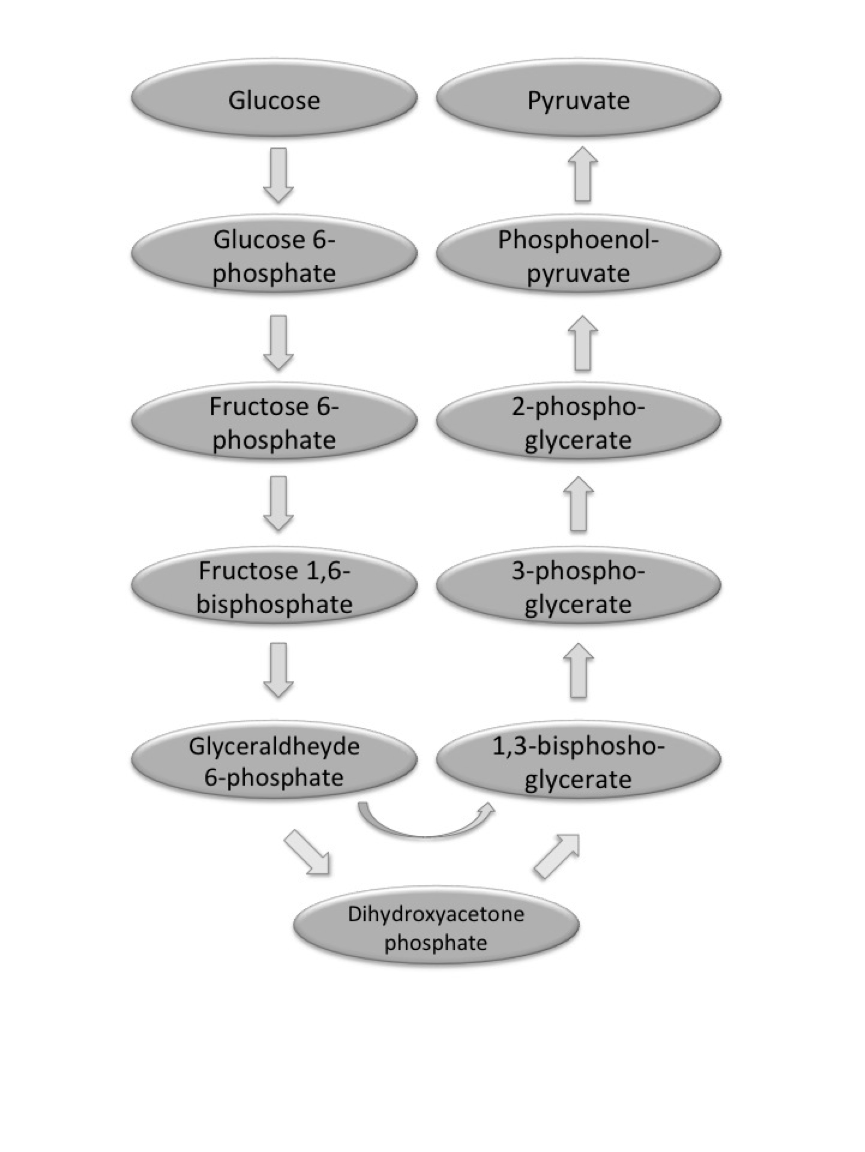}
\caption{A standard view of glycolysis.}
\label{fig:glyco}
\end{figure}

The MetaboX library also outputs a reactants subnetwork which only contains the nodes and the edges of the input metabolites. In this case study, the subnetwork will result in just few nodes and edges (3 nodes and 2 edges) because the compounds choosen from glycolysis represent a subset of the glycolysis pathway (map00010) shown in Fig. \ref{fig:glyco}. Therefore, each metabolite is both a reactant and a product of its neighbors in the pathway. The networks built with the MetaboX library are not pathway representation of the input metabolites.
A classic pathway view in the form of substrate-product flow is provided by many databases of metabolic data (KEGG, MetaCyc, MetaboAnalyst), and it is not implemented in the MetaboX library. Instead, we mainly focused our efforts to highlight other information, like the relationships between metabolites of the same reaction, that until now have never been shown in an automated way. In the case of glycolysis, therefore, the compounds are disconnected between them in the reactants network built only with the input metabolites, because, being each one the precursor of the other, they will never be reagents of the same reaction.

Another example we provide is the construction of a unipartite and bipartite enzymes network. In the first case, shown in Fig. \ref{fig:ec_unipartite}, each node of the network is an enzyme, and two of them are connected if they share at least one metabolite in the reactions they catalyze, with the constraint that the substrate of one enzyme is the product of another enzyme. We added this constraint to allow the user to easily have a view of the substrate-product flow, looking at enzymes instead of metabolites. In this way we were able to build a unipartite enzymes network with 297 nodes and 7705 interactions.
Finally, we built a bipartite enzyme-metabolite network, shown in Fig. \ref{fig:ec_bipartite}. As already mentioned, this network consists of two sets of nodes: metabolites and enzymes. Nodes are connected alternatively, that is a metabolite to an enzyme and vice versa. Connections between two metabolites or two enzymes are not possible. As a resulting network, we were able to find 342 nodes (11 metabolites and 331 enzymes) and 393 interactions. Looking at this network, we can easily identify the hubs (namely glucose, pyruvate, glyceraldehyde 3-phosphate and phosphoenolpyruvate) and all the enzymes related to them.

To change something in these hubs, the variables (enzymes) to be considered are really a lot, and the design of a subsequent experiment in metabolic engineering would be too complicate. The better solution is, instead, to focus attention on the compounds with few connections, in order to limit the analysis to few enzymes and to decrease the complexity of further analyses. Anyhow, we strongly believe that this type of network significantly simplifies the work of those who analyze metabolic pathways to understand metabolic disorders, to connect disease to enzyme defects, to design successful metabolic engineering strategies. MetaboNetworks provides analyses for specific organisms whereas in the MetaboX library we do not account for this feature, considering all available reactions in the first place. In such a way, a user can plan analyses pipelines or methodologies from a compound-wise point of view, and not an organism-wise point of view. For instance, in soil remediation from copper, MetaboX provides access to all known reactions containing copper. A user can then find which organisms use copper within their metabolic pathways.

In the enzyme API call, KEGG provides the \emph{GENES} attribute containing a list of genes where that particular enzyme is involved. Each one of those genes is specific for an organism, enabling the user to filter organism specific reactions.
In conclusion, if we start without any initial information about compounds concentration and we look at a general network, for sure the highly connected nodes are fundamental in the metabolism, and changes in these nodes would have probably led to the death of the organism. On the contrary, if we start from experimental data, it might be useful to correlate increases or decreases of the concentration of some compounds to a particular disease or to a particular disorder. Therefore highlighting compounds within a network should be useful for designing any strategy aimed at clarifying the mode of occurrence of the disease, extracting from that network information like mumber of edges, enzymes, reactions, etc. The MetaboX library is a suitable tool created to solve both issues: a first preliminary view and a second in-depth analysis.

\section{Conclusions}
\label{sec:conc}
In this paper we describe the MetaboX library, a framework to build metabolic networks using information gathered from KEGG database. The advantages of using such a library are: \textit{(i)} the possibility to gather information from KEGG using a collection of KEGG IDs. \textit{(ii)} the possibility to build a representation of the metabolic processes that can highlight how changes in metabolites or enzymes might affect other processes. \textit{(iii)} the possibility to export the networks to other formats for visualization and analysis with standard software, such as Cytoscape, NetworkX, D3JS or D3py.
Because of its extensibility, the MetaboX library may add support to other fields as in the construction of protein-protein interaction (PPI) networks for performing different topological and functional analyses \cite{ankush}. In this case, the MetaboX library should use a resource provider that stores information about interactions between proteins such as STRING \cite{string}. An organism filter can be implemented in the MetaboX library, as used in MetaboNetworks, in order to select specific reactions and build a sub network that enables an organism-wise network. This can be achieved building an organism list for each enzyme of each processed reaction.

The library has been built with the possibility to extend data information gathering, such as downloading these from databases other than KEGG or merge information collected from multiple databases.
Indeed, in order to make the network construction as complete as possible, the MetaboX library will implement a merge process among different resource providers. As KEGG information is limited, it makes sense to gather data about metabolites, reactions, pathways and enzymes from other databases like MetaCyc.

The MetaboX library is the starting point of a three layer project involving a web service and a web application. As stated in (http://www.w3.org/TR/ws-arch/), ''Web services provide a standard means of interoperating between different software applications, running on a variety of platforms and/or frameworks.''. Following this vision, MetaboX library offers a framework to work with metabolic networks. On top of that, we will develop a web service to expose core functionalities to the web. The MetaboX webservice will work in a RESTful fashion, providing APIs to retrieve resources information, network construction options and job submission. Moreover, we will implement alternative ways to make data persistent, such as database storage.

\section*{Acknowledgements} 
This work has been partially funded by MIUR projects\\PON02\_00619 and Italian Flagship project \textit{Interomics}. 
We wish to thank GTP Lab (Genomic, Transcriptomin and Proteomic Laboratory) researchers from ICAR CNR and Dr. Maria Brigida Ferraro for helpful discussion during the early stages of this project and for testing the MetaboX library giving an important feedback for improvements.

\clearpage

\lstset{ %
language=PHP,                
basicstyle=\ttfamily\footnotesize,    
numbers=left,                   
numberstyle=\ttfamily\footnotesize,      
stepnumber=1,                   
numbersep=5pt,                  
backgroundcolor=\color{white},  
showspaces=false,               
showstringspaces=false,         
showtabs=false,                 
frame=single,           
tabsize=2,          
captionpos=b,           
breaklines=true,        
breakatwhitespace=false,    
escapeinside={\%*}{*)}          
}
\begin{lstlisting}[caption=Loading Metabolites and Reactions metadata from KEGG,label={lst:resourcesloader}]
// Retrieve and collect compound information
foreach( $compounds as $compound ){
	$_cpd_id = trim($compound);	
	$cpd_loader = new MetaboX\Resource\Loader\Compound($_cpd_id, $cpdLoaderConfig);
	$_compounds[$_cpd_id] = $cpd_loader->load();
}

// Retrieve and collect reactions information
foreach($_compounds as $id => $compound){
	$rn_list = $compound->reactionIdCollection;
	
	if( $rn_list ){
		foreach( $rn_list as $rn ){
			$_rn_id = trim($rn);			
			$rn_loader = new MetaboX\Resource\Loader\Reaction($_rn_id, $rnLoaderConfig);
			$_reactions[$_rn_id] = $rn_loader->load();
		}	
	}
}

// Create reactants graph
$_graph = new MetaboX\Graph\ReactantsGraph($_reactions);
$_graph->build($compounds);
\end{lstlisting}

\begin{lstlisting}[caption=Loading Enzymes metadata from KEGG,label={lst:enzymeloader}]
// Retrieve and collect reactions information
foreach($_compounds as $id => $compound){
	$ec_list = $compound->enzymeIdCollection;
	
	if( $ec_list ){
		foreach( $ec_list as $ec ){
			$_ec_id = trim($ec);
			$ec_loader = new MetaboX\Resource\Loader\Enzyme($_ec_id, $ecLoaderConfig);
			$_enzymes[$_ec_id] = $ec_loader->load();
		}	
	}
}
\end{lstlisting}


\end{document}